\documentclass[a4paper,fleqn,usenatbib]{mnras}
\usepackage[T1]{fontenc}
\usepackage{ae,aecompl}
\usepackage{graphicx}
\usepackage{amsmath}
\usepackage{amssymb}

\begin{document}
%%%%%%%%%%%%%%%%%%%%%%%%%%%%%%%%%%%%%%%%\par
%\par
%%%%%%%%%%%%%%%%%%%%%%%%%%%%%%%%%%%%%%%%\par
%\par
%\par
%\setlength \parindent{0pt}

\title[Gas phase formation of deuterated formamide]
%{Gas phase formation of the prebiotic molecule formamide: new quantum
%  computations of its deuteration}
{New quantum chemical computations of formamide deuteration support a gas-phase
  formation of this prebiotic molecule}

% The list of authors, and the short list which is used in the headers.\par
% If you need two or more lines of authors, add an extra line using \\newauthor\par
\author[Skouteris et al.]{
D. Skouteris,$^{1}$\thanks{E-mail: dimitrios.skouteris@sns.it} 
F. Vazart,$^{1}$\thanks{E-mail: fanny.vazart@sns.it}
C. Ceccarelli$^{2,5}$\thanks{E-mail: cecilia.ceccarelli@univ-grenoble-alpes.fr} 
 N. Balucani$^{3,5}$\thanks{E-mail: nadia.balucani@unipg.it}
 C. Puzzarini$^{4,5}$\thanks{E-mail: cristina.puzzarini@unibo.it}
V. Barone$^{1}$\thanks{E-mail: vincenzo.barone@sns.it}
\\
% List of institutions\par
$^{1}$Scuola Normale Superiore, Piazza dei Cavalieri 7, 56126 Pisa, Italy\\
$^{2}$ Univ. Grenoble Alpes, CNRS, IPAG, F-38000 Grenoble, France\\
$^{3}$Dipartimento di Chimica, Biologia e Biotecnologie, Universit\`a degli Studi di Perugia, Via Elce di Sotto 8, 06123 Perugia, Italy\\
$^{4}$Dipartimento di Chimica "Giacomo Ciamician", Universit\`a di Bologna, Via F. Selmi 2, 40126 Bologna, Italy\\
$^{5}$INAF - Osservatorio Astrofisico di Arcetri, 50125, Firenze, Italy
}
%    \date{Received - ; accepted -}

\pagerange{\pageref{firstpage}--\pageref{lastpage}} \pubyear{2015}

\maketitle

\label{firstpage}

\begin{abstract} 
  Based on recent work, formamide might be a potentially very
  important molecule in the emergence of terrestrial life. Although
  detected in the interstellar medium for decades, its formation route
  is still debated, whether in the gas phase or on the dust grain
  surfaces. Molecular deuteration has proven to be, in other cases, an
  efficient way to identify how a molecule is synthesised. For
  formamide, new published observations towards the IRAS16293-2422 B
  hot corino show that its three deuterated forms have all the same
  deuteration ratio, 2--5\%, and that this is a factor 3--8 smaller
  than that measured for H$_2$CO towards the IRAS16293-2422
  protostar. Following a previous work on the gas-phase formamide
  formation via the reaction NH$_2$ + H$_2$CO $\to$ HCONH$_2$ + H, we
  present here new calculations of the rate coefficients for the
  production of monodeuterated formamide through the same reaction,
  starting from monodeuterated NH$_2$ or H$_2$CO.  Some misconceptions
  regarding our previous treatment of the reaction are also cleared
  up. The results of the new computations show that, at the 100 K
  temperature of the hot corino, the rate of deuteration of the three
  forms is the same, within 20\%. On the contrary, the reaction
  between non-deuterated species proceeds three times faster than
  that with deuterated ones. These results confirm that a gas-phase route for the 
formation of formamide is perfectly in agreement with the available observations.
\end{abstract}

% Select between one and six entries from the list of approved keywords.\par
% Don't make up new ones.\par

%%%%%%%%%%%%%%%%%%%%%%%%%%%%%%%%%%%%%%%%%%%%%%%%%%\par

%%%%%%%%%%%%%%%%% BODY OF PAPER %%%%%%%%%%%%%%%%%%\par

\section{Introduction}
Among the zoo of detected interstellar molecules formamide (HCONH$_2$)
has a peculiar role in prebiotic chemistry because it can generate both genetic and
metabolic molecules (e.g. Saladino et
al. 2012). It has been speculated that
formamide could have been brought in large quantities on Earth from
the interstellar/planetary dust and comets that rained on primitive
Earth (e.g. Ferus et al. 2014). Recent
observations in star forming
regions (e.g. Lopez-Sepulcre et al. 2015) and, in particular, in regions that will eventually form stars
like the Sun and planetary systems like the Solar System (e.g. Kahane
et al. 2013; Mendoza et al. 2014) are in favour of this scenario. Kahane et al. (2013) have also found that the relative abundances of formamide and water,
[HCONH$_2$]/[H$_2$O], in the solar type protostar IRAS16293-2422
(hereinafter IRAS16293) and in the coma of the comet Hale-Bopp are
very similar. Furthermore, Jaber et al. (2014) have found that
formamide abundance shows an increase by more than two orders of magnitude in the
interior of IRAS16293, in the so-called hot corino region. There, the
formamide abundance is $\sim 6\times 10^{-10}$, equivalent to a mass
of more than ten Mediterranean seas.

One peculiar aspect of the solar type star forming regions is the
largely enhanced deuterium fractionation observed in H-bearing
molecules (e.g. Ceccarelli et al. 2014). Indeed, the deuteration of
molecules formed in the first cold prestellar phase can be several
orders of magnitude larger than the statistical value based on the
elemental D/H ratio ($1.5 \times 10^{-5}$; Linsky 2007). Emblematic
examples are those of formaldehyde and methanol, where the doubly (and
even triply, for methanol) deuterated forms have been detected with
D-bearing/H-bearing ratios up to 30 (and 5) per cent (e.g. Ceccarelli
et al. 1998; Parise et al. 2003, 2006). Interestingly, the deuteration
ratio in different molecules might provide the temporal sequence of
their formation and, possibly, information on their synthetic route
(e.g. Caselli \& Ceccarelli 2012; Ceccarelli et al. 2014).

Very recently, Coutens et al. (2016; hereinafter CJW2016) have
reported the detection of deuterated formamide in the solar-type
protostar IRAS16293 (see Jorgensen et al. 2016 and Jaber Al-Edhari et
al. 2017, for a recent description of the source).  In their work,
CJW2016 detected the three forms of deuterated formamide: trans- and
cis- HCONHD, and DCONH$_2$. They found that they have approximately
the same deuteration ratio, 2-5 \%. Note that the latter observations
have been obtained with the ALMA interferometer and that they refer to
the hot corino region only.  CJW2016 noticed that the
fact that the three forms have a similar deuteration ratio provides
constraints on the synthesis of formamide.  Different routes have been
suggested in the literature: the  gas-phase reaction NH$_2$ + H$_2$CO (Barone et al. 2015;
Vazart et al. 2016), grain-surface radical recombination
(e.g. Fedoseev et al. 2016), synthesis dominated by UV or ion
irradiation of ices containing various species, like methanol
(e.g. Jones et al. 2011, Kanuchova et al. 2016).  CJW2016 noted that
the relatively small percentage of deuterated formamide with respect
to formaldehyde might suggest a preference for synthesis on ice, as
gas phase routes should mostly retain the deuteration ratio of parent
molecules. 

However, this is not necessarily true.
First of all, even though all the isotopologues 
are characterized by the same electronic energies, 
the reaction rate coefficients for the isotopic variants can be influenced by factors such as zero point energies (ZPE) and densities of states. 
In addition, the 
reaction mechanism controls whether D or H displacement is
favored in gas phase reactions involving partially deuterated species. In particular, the release of D over H can 
be favored in direct abstraction processes, while the release of H over D is favored in indirect reactions involving partially deuterated species. Also, the details of the reactive potential
energy surface can change these common-sense rules (Skouteris et
al. 1999).

In this letter, we present new computations on the gas-phase reaction
between partially deuterated amidogen (NH$_2$) or partially deuterated
formaldehyde (H$_2$CO) and discuss the implications for the above
mentioned astronomical observations.

\section{Gas-phase formation of formamide}
\begin{figure}
  \centering
   \includegraphics[angle=0,width=7cm]{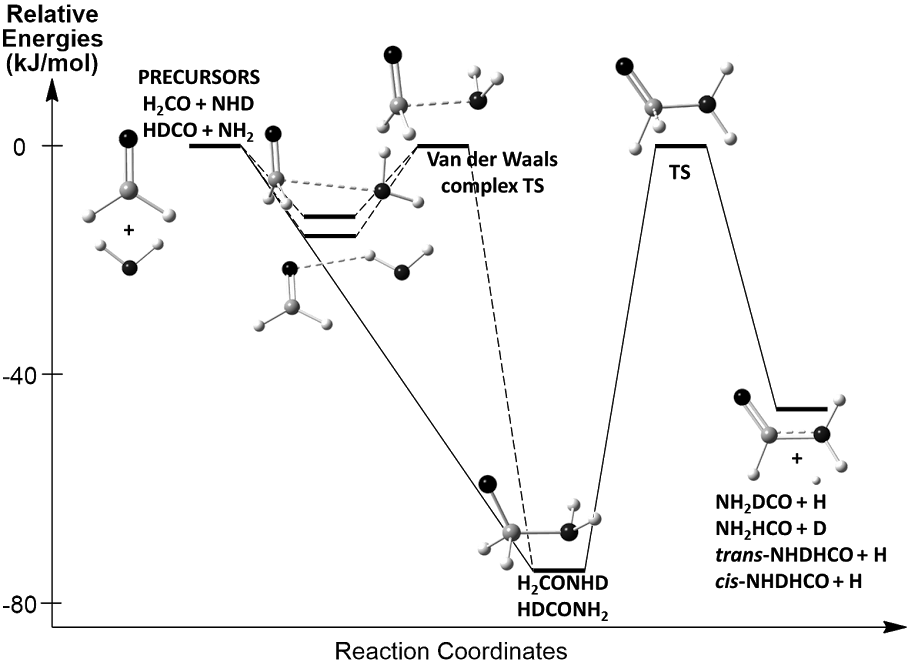}
   \caption{Complete reaction path for the H$_2$CO + NHD $\rightarrow$
     NHDHCO + H and HDCO + NH$_2$ $\rightarrow$ NH$_2$DCO + H processes.}
  \label{fig:path}
\end{figure}

The occurrence of the NH$_2$ + H$_2$CO reaction as a source of formamide
was first suggested by Kahane et al. (2013) but no measurements or
computations were available at the time. Later, Barone et al. (2015)
carried out accurate electronic structure calculations coupled with
capture theory and RRKM kinetic calculations to estimate the rate
coefficient. They found that the reaction can be very fast also in the
cold interstellar conditions. This result was later confirmed by even
more accurate computations by Vazart et al. (2016).  According to
these first studies, the reaction proceeds with the formation of a
bound intermediate and, therefore, emission of H over D should be
favored. Interestingly, before the formation of the addition
intermediate, a van der Waals complex is formed and a transition state
has to be surmounted to form the addition intermediate (see
Fig. \ref{fig:path}). A second transition state connects the bound
intermediate to the final products.  It should be noted that, for this
particular case, the vicinity of the energy levels of the transition
states and the reactant asymptotes is a real computational challenge
because the status of the transition states as "emerged" or
"submerged" can fall within the uncertainty of the calculations (see
Fig. \ref{fig:path}).  In Barone et al. (2015) and Vazart et
al. (2016) the presence of the van der Waals complex has been
deliberately neglected (see below).  In a recently published
theoretical paper on the formation of formamide on ice, a much smaller
estimate of the rate coefficient was reported for the NH$_2$ + H$_2$CO reaction
(Song \& Kastner 2016) because the energy level of the transition
state connecting the van der Waals complex to the addition
intermediate with the inclusion of its ZPE was
found to be significantly above the reactant asymptote.  In this
paper, we clarify further the choice of neglecting the formation of
the van der Waals complex. In addition, since the substitution of an H
atom by a D one affects the density of states of all intermediates and
transition states, as well as the tunnelling efficiency, a thorough
investigation of the NHD + H$_2$CO and NH$_2$ + HDCO reactions is
presented here to derive the specific rate coefficients and to assess
whether they are compatible or not with a formamide formation in the
gas-phase through the above reactions.

In particular, we consider the following isotopologues:
\begin{itemize}
\begin{small}
\item[(1)]{NH$_2$+HDCO $\rightarrow$ DCONH$_2$+H, $\Delta H^\circ_0=-47.29$ kJ mol$^{-1}$} 
\item[(2)]{NH$_2$+HDCO $\rightarrow$ HCONH$_2$+D, $\Delta H^\circ_0 = -39.07$ kJ mol$^{-1}$}
\item[(3)]{NHD+H$_2$CO $\rightarrow$ t-HCONHD+H, $\Delta H^\circ_0 = -48.49$ kJ mol$^{-1}$}
\item[(4)]{NHD+H$_2$CO $\rightarrow$ c-HCONHD+H, $\Delta H^\circ_0 = -48.50$ kJ mol$^{-1}$}
\end{small}
\end{itemize}
The exothermicities given above have been computed as described in Sec. 3.
 For each of the two monodeuterated versions of
the reactants, there are two possible products to consider. In the case
of  HDCO, both normal and C-deuterated formamide
is formed.  In the case of  NHD, HCONHD can be formed in the cis or trans isomers (hindered rotation around the C-N
bond).
%%%%%%%%%%%%%%%%%%%%%%%%%%%%%%%%%%%%%%%%%%%%%%%%%%%%%\par
\section{Computational details and results}

The main body of the calculations has been performed with a development version of the Gaussian suite of programs (Frisch et
al. 2015).  The computations have been carried out using the double-hybrid
B2PLYP functional (Grimme 2006), in conjunction with the m-aug-cc-pVTZ
basis set (Papajak et al. 2009, Dunning 1989), where d functions on
hydrogens have been removed.  Semiempirical dispersion contributions
have also been included into DFT computations by means of the D3 model of
Grimme, leading to the B2PLYP-D3 functional (Goerigk et al. 2011, Grimme
et al. 2011).  Full geometry optimizations, and calculations of frequencies and rotational constants, have been performed for all
minima and transition states of all isotopomers, also checking the nature of the obtained
structures by diagonalizing their  Hessians that have been used also to evaluate ZPEs at the harmonic level. Additional calculations
based on coupled cluster methods have been performed in order to reevaluate
the energies of all species.  The coupled-cluster singles and doubles
approximation augmented by a perturbative treatment of triple
excitations (CCSD(T), Raghavachari et al. 1989) has been employed in
conjunction with extrapolation to the complete basis set limit and
inclusion of core-correlation effects (CCSD(T)/CBS+CV).  At this level
of calculations, the energy level of the van der Waals transition
state is slightly above the energy of the reactants' asymptote (see
Table 1). However, when a full treatment of triple (fT) and quadruple
excitations (fQ) has also been included, thus leading to the CCSD(T)/CBS+CV+fT and CCSD(T)/CBS+CV+fT+fQ approaches, the barrier issuing
from this transition state significantly decreases (3.6 kJ/mol drops
to 2.05 kJ/mol with full-T and to 1.67 kJ/mol with full-T and full-Q).

 As in previous work (Barone et al. 2015), we have used the
results of the electronic structure calculations to derive the capture
theory rate constant for the formation of the initial bound
intermediate (slightly dependent on the reduced mass).  Subsequently, using the RRKM scheme as in previous
work (Balucani et al. (2012), Leonori et al. (2013), Vazart et al. (2015a)), we have calculated
rate constants (as a function of energy) for the elimination of a
 H or D atom leading to formation of formamide.  We
have also calculated (using detailed balance) the energy-dependent rate constant for
back-dissociation of the intermediate to the reactants as done in Vazart et al. (2015b) and Skouteris et al. (2015).
The master equation has been solved as a function of energy to determine
the bimolecular rate constant for formamide formation and its
isotopologues. Finally, Boltzmann averaging has been carried out to
determine rate constants as a function of temperature.  Each rate constant has been fitted to the rate law: 
%\par
\begin{equation}
\begin{small}
k(T) = A \times \left( T / 300K \right)^{\beta} \times e^{- \gamma / T}
\end{small}
\end{equation}
The results are listed in Table \ref{tab:results} together with those of the 
undeuterated case (from Vazart et al. 2016) for comparison.
\begin{table*}
  \centering
  \begin{tabular}{lccc}
    \hline
    Species & Electronic energy & ZPE corrected (X=H,Y=D) & ZPE corrected (X=D,Y=H)\\
    \hline
    NHX + HYCO & -170.367866 (0.00) & -170.324839 (0.00) & -170.324535 (0.00) \\
    HYCONHX & -170.396176 (-74.33) & -170.344028 (-50.38) & -170.344203 (-51.64)\\
    Transition state & -170.368405 (-1.41)  & -170.323737 (2.89) (DCONH$_2$) & -170.323873 (1.74) (cis)\\
    & &  -170.321363 (9.13) (HCONH$_2$) &  -170.323877 (1.73) (trans)\\
    Products & -170.385371 (-45.96) & -170.342852 (-47.29) (DCONH$_2$) & -170.343008 (-48.50) (cis)\\
   &  & -170.339721 (-39.07) (HCONH$_2$) &  -170.343005 (-48.49) (trans)\\
    \hline
 \end{tabular}
 \caption{Summary of the energy of each species involved in the deuterated versions of the 
   NH$_2$ + H$_2$CO reaction. Each energy is given in hartree and in parentheses is given its relative energy with 
   respect to the reactants in kJ mol$^{-1}$. The first column is the purely electronic (isotope independent) 
   energy, while the others include zero point contributions.} 
  \label{tab:energies}
\end{table*}

\begin{figure}
  \centering
   \includegraphics[angle=0,width=8cm]{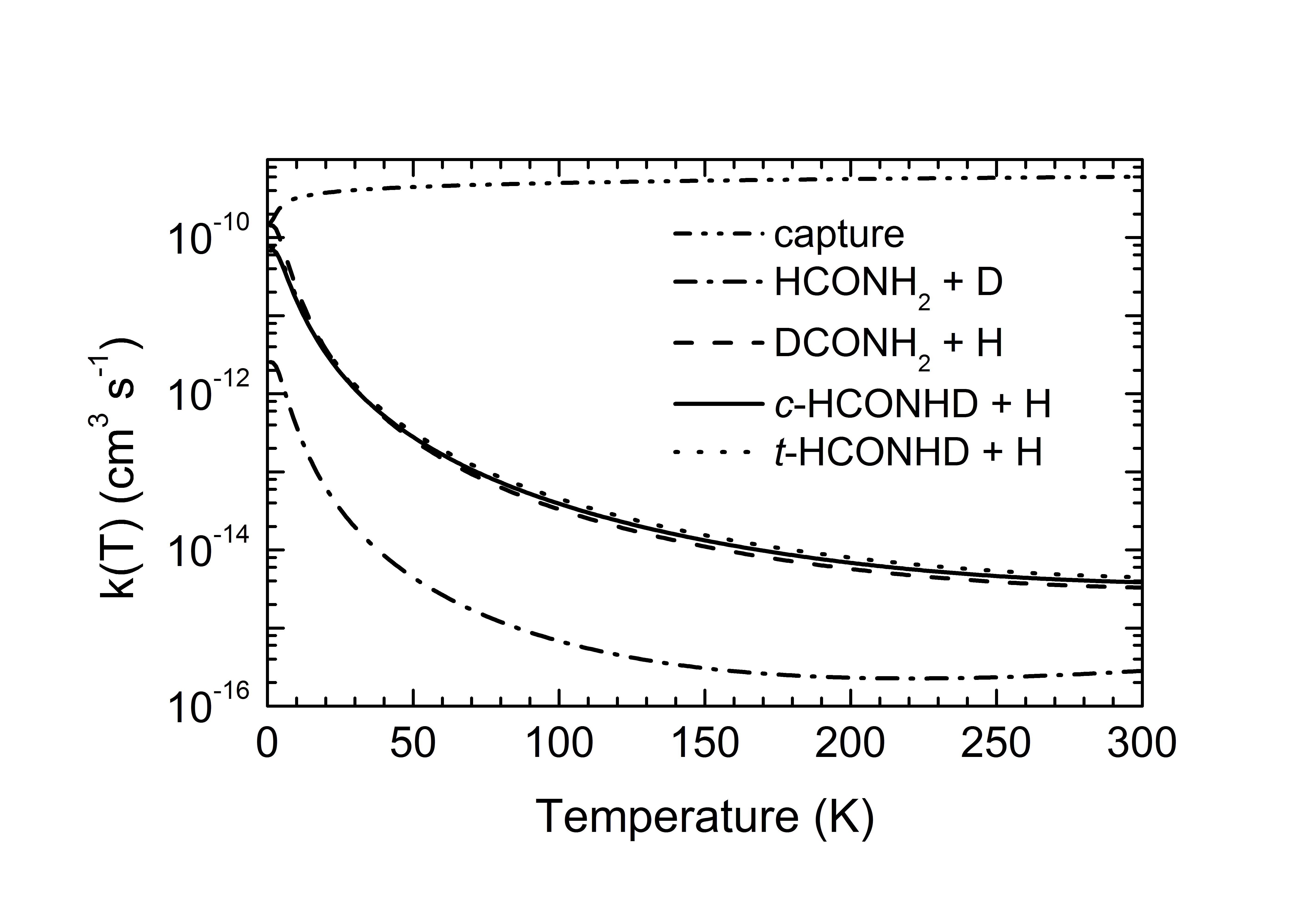}
   \caption{Canonical rate constants for the four different
     deuteration reactions (1)-(4) of Table \ref{tab:results} as a function of
     temperature. The capture rate constant is also shown for
     comparison.}
  \label{fig:curves}
\end{figure}
%%%%%%%%%%%%%%%%%%%%%%%%%%%%%%%%%%%%%%%%%%%%%%%%%%%%%\par
\section{Results and discussion}

\begin{table*}
  \centering
  \begin{tabular}{lcccccc}
    \hline
    Reaction & A & $\beta$ & $\gamma$ & k(T=10 K) & k(T=60 K) & k(T=100 K)\\
                & ($\times 10^{-16}$cm$^3$s$^{-1}$) &  & (K) & \multicolumn{3}{c}{($\times 10^{-13}$cm$^3$s$^{-1}$)} \\
    \hline
    (1) NH$_2$ + HDCO $\rightarrow$ DCONH$_2$ + H & 20.7 & -2.75 & 4.34 & 153 & 1.60    & 0.41\\
    (2) NH$_2$ + HDCO $\rightarrow$ HCONH$_2$ + D & 1.08 & -2.15 & 0.96 & 1.49 & 0.03    & 0.01\\
    (3) NHD + H$_2$CO $\rightarrow$ trans-HCONHD + H & 30.7 & -2.63 & 5.05 & 141 & 1.93 & 0.52\\
    (4) NHD + H$_2$CO $\rightarrow$ cis-HCONHD + H & 26.2 & -2.64 & 5.04 & 126 & 1.69    & 0.45\\
    (5) NH$_2$ + H$_2$CO $\rightarrow$ HCONH$_2$ + H & 77.9 & -2.56 & 4.88 & 287 & 4.41 & 1.23\\
    \hline
 \end{tabular}
 \caption{Summary of the rate coefficients of the reactions involving
   deuterated formamide. The undeuterated case is also reported
   (Vazart et al. 2016). 
   The rates are given in the usual form, $A \times
   \left( \frac{T}{300K} \right)^{\beta} \times \exp (-\gamma /
   T)$. The last three columns report the values of the reaction rate
   at 10, 60 and 100 K, respectively,  namely the
   temperature of a cold molecular cloud, of a shocked gas (e.g. Mendoza et
   al. 2014), and of the IRAS16293-2422 B hot corino where
   formamide has been detected (Jaber et al. 2014;
   Coutens et al. 2016).}
  \label{tab:results}
\end{table*}

The interaction between the reactants starts with the formation of a 
shallow van der Waals complex and the subsequent formation of a bound intermediate.  This can then
eliminate an H atom to form formamide or back-dissociate to the
reactants.  The electronic potential energy surface is the same for
all isotopomers, but the energy levels, vibrational frequencies,
rotational constants and zero point energies depend on the masses involved. This can change the relative densities of states and
the resulting reaction rate constants.  
As already mentioned, Song \& Kastner (2016) disputed the gas-phase
formation of formamide through the NH$_2$ + H$_2$CO reaction, arguing
that the transition state connecting the van der Waals complex to the intermediate is too high when including the ZPE (from +2.7 kJ mol$^{-1}$ to 
17.8 kJ mol$^{-1}$ with ZPE at the  UCCSD(T)-
F12/cc-pVTZ-F12 level of calculations) and
the reaction rate coefficient very small (of the order of
$10^{-22}$ cm$^3$ s$^{-1}$).  Against this claim, it must be stressed
that: $i)$ Even though the calculated energy of the transition state
    leading from the van der Waals complex to the bound intermediate,
    as results from the coupled cluster calculations, is higher
    than the reactant asymptote, its energy substantially drops
    by including higher excitation orders in the cluster operator (see values in the previous section).
    Extrapolating to the full configuration interaction limit, the electronic energy of this transition state
    drops slightly below the reactant level (Vazart et al. 2016) in line with the variational principle of approximating from above.
 $ii)$ The use of the ZPE correction for the van der Waals complex and its transition state is not warranted. In an electronic
    calculation the ZPE is derived from local considerations on the
    potential minimum (harmonic or anharmonic).  However, the
    three new vibrational modes in the van der Waals complex consist
    of a very loose stretching mode (where even a perturbative
    approach including anharmonicity would tend to grossly overestimate the
    frequency) and two loose bending modes that 
    constitute almost free rotations.
    Under these conditions, in order to derive the exact ground state
    level one would have to exactly solve the
    Schr\"odinger equation on the overall potential. We believe that
    it is a much more realistic approach to neglect what emerges as
    "the  ZPE".
Taking into account the two points above,  it is much more reasonable
to omit the van der Waals complex from the reaction scheme rather than largely overestimate its role as done by Song \& Kastner (2016).
 Kinetics experiments should be performed at very low temperatures in a CRESU apparatus to verify our suggestion.

Once formed, the
bound addition intermediate has two possible fates: either {\it a) }   eliminate an H/D atom and form a formamide molecule or {\it b)}
 dissociate,  returning to the original reactants. The relative rates of the two competing steps are
important in determining overall rate constants.  It
is very important to stress that back-dissociation of the addition
intermediate is in competition with the reactive event and its
consideration in the calculations is essential to obtain meaningful
rate coefficients as it reduces a lot their real values.  The practice of giving the capture rate coefficient as
the global rate coefficient is therefore wrong and should be avoided.
Fig. \ref{fig:curves} illustrates this point, showing the canonical
rate constants for all deuteration reactions compared to the capture
rate constant. It can be seen that all reaction rate constants
decrease at high enough energies while the overall capture rate
constant increases. This is because of the rapidly increasing
back-dissociation rate which renders simple capture rate constants
completely unrealistic as estimates of reaction rate coefficients.
Concerning the  temperature trend, we note that at low energies (temperatures), the reaction is promoted by
tunnelling while the rate of back-dissociation is negligible (as a result of the relative translation density of states
tending to zero, in accordance with the detailed balance
principle). As a result, reaction predominates. On the contrary, 
at higher temperatures back-dissociation predominates because the relative translation density of states increases more rapidly than the one of the transition state 
and tunnelling is no longer sufficient.

The main focus of these new calculations concerns the rate coefficients and product branching ratios of the partially deuterated species. The most relevant results can so be summarized: 
$i)$ the reactions involving the exit of an H atom have essentially equal rates, which are lower by a factor of 2-4 than the rate of the undeuterated reaction. 
For the NHD reaction, the constant capture incoming flux is split between two distinct product channels accounting for a factor of 2 (that is the observed difference at 10 K). 
The larger factor observed at higher temperatures and the reduced k(T) for the NH$_2$ + HDCO $\to$ DCONH$_2$ + H rate are, instead, direct consequences of the prevailing back-dissociation; 
$ii)$ the reaction NH$_2$ + HDCO mainly produces DCONH$_2$, with
    only 1$-$2\% going into HCONH$_2$. The reason for this is
    twofold: the lighter H atom tunnels more easily through
    the barrier towards the products by a factor of around 50 and the vibrational ZPE of the transition
    state is lower when an H rather than a D atom has started dissociating. 

In conclusion, even though the partially deuterated formamide products retain the degree of deuteration of the parent molecules,
the three isotopic variants of the NH$_2$ + H$_2$CO reactions are not characterized by the same rate coefficient. In particular,
since the fastest reaction is the one with undeuterated reactants, the expected value of the degree of formamide deuteration at 100 K is a factor of 3 lower
than that of the parent molecules H$_2$CO or NH$_2$. 

\section{Comparison with astronomical observations and conclusions}

So far, deuterated formamide has only been detected towards the hot
corino of IRAS16293-2422 B by CJW2016. They found that the
DCONH$_2$/HCONH$_2$ and HCONHD/HCONH$_2$ abundance ratios are similar
and around 0.02--0.05. Based on this and on the fact that these ratios
are smaller than the measured HDCO/H$_2$CO abundance ratio, CJW2016
favoured a grain-surface formation of NH$_2$CHO. With the new
computations reported in Table \ref{tab:results}, we can now
quantitatively discuss this issue. For that, we will consider the
rates computed for a gas temperature of 100 K, namely the temperature
in the IRAS16293-2422 B hot corino (e.g. Jaber et al. 2014).
\begin{enumerate}
\item {\it DCONH$_2$/HCONH$_2$ and HCONHD/HCONH$_2$ abundance ratios
    smaller than HDCO/H$_2$CO:} The values in Table \ref{tab:results}
  show that there is a factor three difference in the rates of
  formation of H- against D- formamide. Therefore, the deuteration of
  formamide synthesised in the gas-phase will be three times smaller
  than that of the mother molecules NH$_2$ and H$_2$CO. No observations of
  NHD/NH$_2$ exist so far, so no constraints can be obtained from
  amidogen. Similarly, no observations exist of the HDCO/H$_2$CO
  abundance ratio in the hot corino of IRAS16293-2422 B. However,
  sigle-dish observations of deuterated formaldehyde towards IRAS16293
  measured HDCO/H$_2$CO$\sim0.15$ (e.g. Ceccarelli et al. 1998;
  Loinard et al. 2000), namely a factor 3--8 larger than
  DCONH$_2$/HCONH$_2$.  Therefore, within the uncertainty of the
  available observations, the hypothesis of formamide gas formation is
  fully consistent with them.  In addition, as correctly pointed out
  by CJW2016, single-dish observations encompass both the hot corino
  and the extended envelope. Ceccarelli et al. (2001) mapped the
  D$_2$CO line emission towards IRAS16293 and showed that the
  D$_2$CO/H$_2$CO abundance ratio is about 0.03. Likewise,
  HDCO/H$_2$CO is also high in the extended envelope and, thus, the
  single-dish measurements are likely largely contaminated by the
  extended envelope.
\item {\it Similar DCONH$_2$/HCONH$_2$ and HCONHD/HCONH$_2$ abundance
  ratios:} The values in Table \ref{tab:results} show that indeed the
  two abundance ratios are similar (0.8-0.9) if the gaseous NHD/NH$_2$
  and HDCO/H$_2$CO are similar. In this respect, then, the gas-phase
  formation of formamide is perfectly consistent with the observed
  values.
\end{enumerate}
We conclude that the available
observations do not allow to rule out a gas-phase formation route for
formamide but, on the contrary, they support it.  New high-spatial
resolution observations of {\it both} NHD and HDCO towards IRAS16293
will be necessary to challenge this conclusion.

 Finally, we would like to note that the same approach could be used to  address the deuteration of other complex organic molecules,
such as methyl formate, for which a debate on gas-phase versus ice-assisted formation routes is currently ongoing (see, for instance, Balucani et al. 2015).
%%%%%%%%%%%%%%%%%%%%%%%%%%%%%%%%%%%%%%%%%%%%%%%%%%%%%\par
\section{Acknowledgements}
CC thanks Audrey Coutens for clarifying some aspects of her
article. NB acknowledges the financial support from the Universit\'e
Grenoble Alpes and the Observatoire de Grenoble. The research leading
to these results has received funding from the European Research
Council under the European Union's Seventh Framework Programme
(FP/2007-2013) / ERC Grant Agreement n. [320951]. CP acknowledges the financial support from 
Italian MIUR under the research funding scheme PRIN 2012 (Project "STAR: Spectroscopic and 
computational Techniques for Astrophysical and atmospheric Research").

%%%%%%%%%%%%%%%%%%%%%%%%%%%%%%%%%%%%%%%%%%%%%%%%%%\par

%%%%%%%%%%%%%%%%%%%% REFERENCES %%%%%%%%%%%%%%%%%%\par

% Don't change these lines\par
\bsp %typesetting comment\par

\label{lastpage}


\begin{thebibliography}{}
%\\bibitem[\\protect\\citeauthoryear\{\}\{\}]\{\} \par

\bibitem[\protect\citeauthoryear{balucani}{2015}]{balucani2015} 
Balucani N., Ceccarelli C., Taquet V. 2015, MNRAS, 449, L16

\bibitem[\protect\citeauthoryear{balucani}{2012}]{balucani2012} 
Balucani N., Skouteris D., Leonori F., Petrucci R., Hamberg M., Geppert W.D., Casavecchia P., Rosi M.,
 2012, J. Phys. Chem. A 116 (43), 10467 

\bibitem[\protect\citeauthoryear{barone}{2015}]{barone2015} 
Barone V., Latouche C., Skouteris D., Vazart F., Balucani
  N., Ceccarelli C., Lefloch B. 2015, MNRAS, 453, L31

\bibitem[\protect\citeauthoryear{Caselli \& Ceccarelli}{2012}]{Caselli2012} 
Caselli P. \& Ceccarelli C., 2012, A\&A Rev, 20, 56

\bibitem[\protect\citeauthoryear{Ceccarelli}{1998}]{ceccarelli1998}  
Ceccarelli C., Castets A., Loinard L., Caux E., Tielens A. 1998, A\&A 338, L43

\bibitem[\protect\citeauthoryear{Ceccarelli}{2001}]{ceccarelli2001}  
Ceccarelli C., Loinard L., Castets A., et al. 2001, A\&A 372, 998

\bibitem[\protect\citeauthoryear{Ceccarelli}{2014}]{ceccarelli2014} 
Ceccarelli C., Caselli P., Bockeleen-Morvan D., et al. 2014,
Protostars and Planets VI, eds. E. Beuther, R.S. Klessen,
C.P. Dullemond, and T. Henning, University of
Arizona Press, p. 859

\bibitem[\protect\citeauthoryear{coutens}{2016}]{coutens2016} 
Coutens A., Jorgensen J. K., van der Wiel M. H. D. et al. 2016, A\&A
590, L6 (CJW2016)

\bibitem[\protect\citeauthoryear{Dunning}{1989}]{Dunning1989} 
Dunning T. H. 1989, J. Chem. Phys. 90, 1007

\bibitem[\protect\citeauthoryear{Ferus et al.}{2014}]{Ferus2014}
Ferus M., Nesvornyc D., Sponer J., et al. 2014 PNAS

\bibitem[\protect\citeauthoryear{Frisch et al.}{2013}]{Frisch2013}
  Frisch, M. J. et al. Gaussian09 GDVH32, 2013, GDVH32
\
\bibitem[\protect\citeauthoryear{Goerigk \&Grimme}{2011}]{GoerigkGrimme2011} 
Goerigk L., Grimme S. 2011, J. Chem. Theory Comput. 7, 291-309

\bibitem[\protect\citeauthoryear{Grimme}{2006}]{Grimme2006} Grimme S. J. 2006 Chem. Phys. 124, 034108

\bibitem[\protect\citeauthoryear{Grimme et al.}{2011}]{Grimme2011}
  Grimme S., Ehrlich S., Goerigk L., 2011, J. Comput. Chem. 32, 1456-1465

\bibitem[\protect\citeauthoryear{Jaber et al.}{2014}]{Jaber2014}
Jaber A., Ceccarelli C., Kahane C., Caux E. 2014, ApJ 791, 29

\bibitem[\protect\citeauthoryear{Jaber et al.}{2017}]{Jaber2017}
Jaber Al-Edhari A., Ceccarelli C., Kahane C. et al. 2017, A\&A, 597, 40 

\bibitem[\protect\citeauthoryear{Jones et al.}{2011}]{Jones2011}
Jones B.M., Bennett C.J., Kaiser R.I., 2016, ApJ, 734, 78

\bibitem[\protect\citeauthoryear{Jorgensen et al.}{2014}]{Jorgensen2016}
Jorgensen J.K., van der Wiel M.H.D., Coutens A., et al. 2016

\bibitem[\protect\citeauthoryear{Kahane et al.}{2013}]{Kahane2013}
Kahane C., Ceccarelli C., Faure A., Caux E., 2013, ApJ, 763, L38

\bibitem[\protect\citeauthoryear{kanuchiva}{2016}]{kanuchova2016} 
Kanuchova G., Urso R., Baratta G. et al. 2016, A\&A 585, 155

\bibitem[\protect\citeauthoryear{leonori}{2013}]{leonori2013} 
Leonori F., Skouteris D., Petrucci R., Casavecchia P., Rosi M., Balucani N.,
 2013, J. Chem. Phys. 138 (2), 024311

\bibitem[\protect\citeauthoryear{Lynsky}{2007}]{linsky2007}
Linsky J. L. (2007) Space Sci. Rev., 130, 367.
 
\bibitem[\protect\citeauthoryear{Lopez-Sepulcre et al.}{2015}]{Lopez-Sepulcre2015}
Lopez-Sepulcre A. et al., 2015, MNRAS 449, 2438

\bibitem[\protect\citeauthoryear{Mendoza et al.}{2014}]{Mendoza2014}
Mendoza E. et al., 2014, MNRAS, 445, 151

\bibitem[\protect\citeauthoryear{Noble et al.}{2015}]{Noble2015}
Noble J. A. et al. 2015, A\&A 576, 91

\bibitem[\protect\citeauthoryear{Papajak et al.}{2009}]{Papajak2009} 
Papajak, E. et al. 2009, J. Chem. Theory Comput. 5, 1197-1202

\bibitem[\protect\citeauthoryear{parise et al.}{2003}]{parise2003}
Parise B., Castets A., Herbst E. et al. 2003, A\&A  410, 897.

\bibitem[\protect\citeauthoryear{parise et al.}{2006}]{parise2006}
Parise B, Ceccarelli C., Tielens A. . et al. 2006, A\&A 453, 949.

\bibitem[\protect\citeauthoryear{Raghavachari et al.}{1989}]{raghavachari1989}
Raghavachari K., Trucks G. W., Pople J. A., Head-Gordon M., 1989, Chem. Phys. Lett. 157, 479.

\bibitem[\protect\citeauthoryear{Saladino et al.}{2012}]{Saladino2012}
Saladino R., Botta G., Pino S., Costanzo G., Di Mauro E., 2012, Chem. Soc.
Rev. 41, 5526.

\bibitem[\protect\citeauthoryear{Skouteris et al.}{2015}]{Skouteris2015}
Skouteris D., Balucani N., Faginas-Lago N., Falcinelli S., Rosi M., 2015, A\&A 
584, A76. 

\bibitem[\protect\citeauthoryear{Skouteris et al.}{1999}]{Skouteris1999}
Skouteris D., Manolopoulos D.E., Bian W., Werner H.-J., Lai L.H., Liu K., 1999, Science, 286 (5445), 1713. 

\bibitem[\protect\citeauthoryear{Song}{2016}]{Song2016} 
Song L. \& K\"astner J., 2016, Phys. Chem. Chem. Phys., 18, 29278.

\bibitem[\protect\citeauthoryear{taquet}{2016}]{taquet2016}
Taquet V., Wistrom E., Charnely S. 2016, ApJ 821, 46

\bibitem[\protect\citeauthoryear{Vazart et al.}{2015}]{Vazart2015a}
  Vazart F., Calderini D., Skouteris D., Latouche C., Barone V., 2015, J. Chem. Theory 
Comput. 11(3), 1165.

\bibitem[\protect\citeauthoryear{Vazart et al.}{2015}]{Vazart2015b}
  Vazart F., Latouche C., Skouteris D., Balucani N., Barone V., 2015, ApJ 810, 2.

\bibitem[\protect\citeauthoryear{Vazart et al.}{2016}]{Vazart2016}
  Vazart F., Calderini D., Puzzarini C. et al., 2016, J. Chem. Theory 
Comput. 12(11), 5385.

\end{thebibliography}
\end{document}